\documentclass[epj]{svjour}

\usepackage{graphicx}
\usepackage{fancyhdr}
\def\makeheadbox{{
 }}
\def\mytitle{Supersymmetric Interpretation of the EGRET Excess in Diffuse Galactic Gamma Rays} 
\def\myauthors{Christian Sander}  
\def\mytype{Contributed Talk}
\def\mysession{Cosmology and Astrophysics}

\begin{document}
\title{Supersymmetric Interpretation of the EGRET Excess in Diffuse Galactic Gamma Rays}
\author{Christian Sander
\thanks{\emph{Email:} christian.sander@cern.ch}%
}                     
%
%
\institute{Institute f\"ur Experimentalphysik, University of Hamburg, Germany
}
%
\date{}
\abstract{Recently it was shown that the excess of diffuse Galactic gamma rays above 1 GeV could be interpreted as a Dark Matter annihilation signal. From the spectral shape of the excess it is possible to determine a range for the allowed WIMP mass which can be used to test the supersymmetric parameter space. It is found that the EGRET excess combined with electroweak and other constraints, e.g. relic denisity or direct detection limits, is fully consistent with the minimal mSUGRA model for scalars in the TeV range and gauginos below 500 GeV.
\PACS{
      {12.60.Jv}{Supersymmetric models}   \and
      {95.35.+d}{Dark matter}
     } 
} 
\maketitle
%

\section{Introduction} \label{intro}
Cold Dark Matter (CDM) makes up 23\% of the energy density of the universe \cite{wmap}. One of the most popular CDM candidates is the neutralino, a stable neutral particle predicted by Supersymmetry. The neutralinos are spin 1/2 Majorana particles, which can annihilate into pairs of Standard Model (SM) particles. A large fraction of the annihilations is expected to go into quark-antiquark pairs. Since the DM particles are strongly non-relativistic, the produced particles are mono-energetic. In a recent paper we showed that the observed excess of diffuse Galactic gamma rays above 1 GeV, measured by the EGRET experiment, has all the properties of the $\pi^0$ decays of such mono-energetic quarks originating from the annihilation of neutralinos with a mass around 60 GeV \cite{us}. It was also shown that a newly determined extragalactic gamma ray background including Dark Matter annihilation (DMA) shows a less significant bump at the same energies \cite{egb}.

It is the purpose of the present paper to see if this intriguing hint of DMA is compatible with Supersymmetry. Here we will concentrate on the Minimal Supersymmetric Model with supergravity inspired symmetry breaking (mSUGRA model). We assume that the EGRET excess originates from the annihilation of the stable, neutral lightest supersymmetric particles, the neutralinos. Their mass is then constrained to be between 50 and 100 GeV from the EGRET data, which strongly constrains the masses from all other SUSY particles, if mass unification at the GUT scale is assumed. The upper limit on the WIMP mass depends strongly on the background model of the Galactic gamma rays. If one uses a conventional galactic model which reproduces the locally measured electron and proton fluxes as well as secondary to primary ratios, one finds an upper limit of $\sim 70$ GeV. If one uses a model which was optimized to fit the gamma ray excess, it is found that the excess is still existent but not as strong as before, which leads to a higher mass limit of $\sim 100$ GeV. It will be shown that combining the EGRET data with other constraints, like the electroweak precision data, Higgs mass limits, chargino limits and relic density leads to a very constrained SUSY mass spectrum with light gauginos and heavy squarks and sleptons \cite{pl}.

\section{Predictions from mSUGRA} \label{susywimp}
The mSUGRA model, i.e. the Minimal Supersymmetric Standard Model (MSSM) with supergravity inspired breaking terms, is characterised by only 5 parameters: $m_0$, $m_{1/2}$, $\tan\beta$, sign($\mu$), $A_0$. Here $m_0$ and $m_{1/2}$ are the common masses for the gauginos and scalars at the GUT scale, which is determined by the unification of the gauge couplings. Gauge unification is still possible with the precisely measured couplings at LEP \cite{bs}. The ratio of the vacuum expectation values of the neutral components of the two Higgs doublets in Supersymmetry is called $\tan\beta$ and $A_0$ is the trilinear coupling at the GUT scale. We only consider the dominant trilinear couplings of the third generation of quarks and leptons and assume also $A_0$ to be unified at the GUT scale. Electroweak symmetry breaking fixes the scale of $\mu$, so only its sign is a free parameter. We use the positive sign, as suggested by the small deviation of the muon anomalous moment.

\begin{figure}
 \begin{center}
  \includegraphics [width=0.50\textwidth,clip]{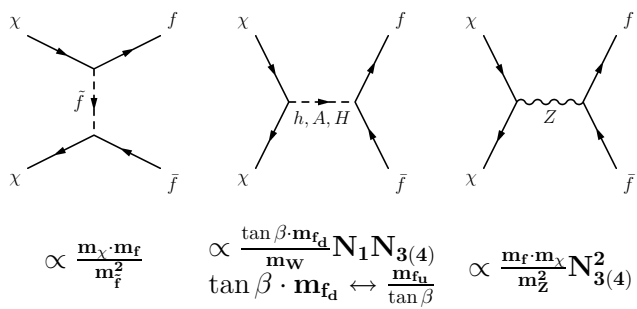}
  \caption[]{The dominant annihilation diagrams for the lightest neutralino, which is a linear combination of the gaugino and Higgsino states: $|\chi_0\rangle =N_1 |B_0\rangle +N_2|W^3_0\rangle +N_3|H_1\rangle +N_4|H_2\rangle$. The dependence of the amplitudes on masses and neutralino mixing parameter $N_i$ has been indicated.} \label{diagrams}
 \end{center}
\end{figure}

The dominant annihilation diagrams of the lightest supersymmetric particle (LSP) neutralino are shown in figure \ref{diagrams}. The cross sections are proportional to the final state fermion mass, which originates either from the Yukawa couplings for the Higgs exchange diagram or from the helicity suppression at the low energies involved in cold DMA. Therefore heavy fermion final states, i.e. third generation quarks and leptons, are expected to be dominant. The Higgs exchange diagram is in addition proportional to $\tan\beta$ for down type quarks and $1/\tan\beta$ for up type quarks, indicating that top quark final states are suppressed for large $\tan\beta$. 

The $W$- and $Z$-final states have usually a much smaller cross section due to the weak couplings involved and are in addition kinematically suppressed for the 60 GeV neutralino mass preferred by the EGRET data.

\begin{figure}
 \begin{center}
  \includegraphics [width=0.40\textwidth,clip]{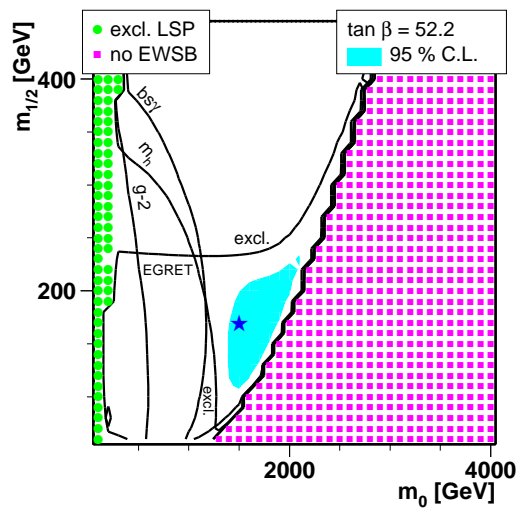}
  \caption[]{The shaded area (blue) indicates the 95\% CL parameter range in the $m_0$-$m_{1/2}$-plane allowed by the EGRET data, if the constraints from electroweak data, a neutral LSP and electroweak symmetry breaking (EWSB) are imposed as well. The individual constraints have been indicated by the lines and dots. The values of $A_0=0$ and $\tan\beta=52.2$ were chosen to be compatible with WMAP for the parameters marked by the star.}\label{msugra1}
 \end{center}
\end{figure}

For the pseudoscalar Higgs and sfermion exchange diagrams the annihilation rate, which is proportional to the cross section multiplied by the relative neutralino velocity, is practically independent of the center of mass energy, but the other diagrams show a strong energy dependence \cite{pl}. This implies that for the present temperature of the universe close to absolute zero the neutralino annihilation is dominated by either sfermion exchange or pseudoscalar Higgs exchange. The sfermion exchange is suppressed for the following reason. The Born mass of the lightest Higgs is below the $Z^0$ mass, but radiative corrections can boost it up to 130 GeV in the minimal mSUGRA model. These corrections depend on the heavy particles coupling to the lightest Higgs, like the top and stop quarks. These scalars have to be sufficiently heavy in order to reach a Higgs mass above 114 GeV, which is the present lower limit from the direct searches at LEP \cite{lephiggs}. Note that this is the limit on the Standard Model Higgs particle, but for heavy scalars the lightest SUSY Higgs particle has very much the properties of the SM Higgs, so the above limit is also valid in our case. For light neutralinos, i.e. small $m_{1/2}$, the Higgs mass limit excludes scalar masses below the TeV range, as indicated in Fig. \ref{msugra1}, by the almost vertical line, labeled $m_h$. In addition, the excluded regions from $Br(B \to X_s\gamma)$ and the anomalous magnetic moment of the muon have been indicated (left from the corresponding lines). The mass spectrum has been calculated with the \texttt{Suspect} program \cite{suspect} while the quantities $\Omega h^2$, $Br(B \to X_s\gamma)$ and $\Delta a^{\rm SUSY}_\mu$ were calculated with \texttt{micrOMEGAs} \cite{micromegas}. For the exclusion limits the following inputs were used: a) $Br(B \to X_s\gamma)=(3.43\pm 0.36)\cdot 10^{-4}$, which is the average from BaBar, CLEO and BELLE \cite{bsgexp}. b) the deviation of the anomalous magnetic moment of the muon $a_\mu$ from the expected value in the Standard Model was taken to be \cite{amu}: $\Delta a_\mu=a_\mu^{\mbox{\scriptsize{exp}}}-a_\mu^{\mbox{\scriptsize{theo}}}=(27\pm 10)\cdot 10^{-10}$.

The constraints on the supersymmetric parameters space are practically independent of $A_0$ due to a coincidence from the constrains from the $Br(B \to X_s\gamma)$ rate and the lower limit on the Higgs mass of 114 GeV \cite{bs}. The region of large $m_0$, for which no electroweak symmetry breaking (EWSB) is possible, has been indicated in Fig. \ref{msugra1} as well as the region of small $m_0$, where the stau would be the lightest SUSY partice, which is excluded, since the DM candidate has to be neutral. The $m_0$ values in the TeV range between the Higgs mass limit and EWSB limit are allowed by all constraints considered sofar.

An independent check that the sfermion exchange diagram is strongly suppressed comes directly from the EGRET data: if the scalars are light, the stau is usually the lightest scalar, in which case the stau exchange in the t-channel (left diagram of Fig. \ref{diagrams}) would be dominant, thus leading to tau final states. The low decay multiplicity of tau leptons leads to a much harder gamma ray spectrum from the hadronic decays, which is excluded by the EGRET data.

\begin{figure}
 \begin{center}
  \includegraphics [width=0.40\textwidth,clip]{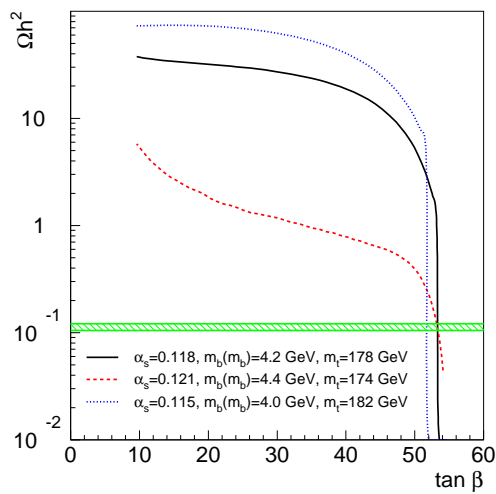}
  \caption[]{$\Omega h^2$ as a function of $\tan\beta$ is plotted for $m_0=1500$ GeV, $m_{1/2}=200$ GeV, $A_0=0$. The dependence on the uncertainty of the SM parameters ($\alpha_s$, $m_b$ and $m_t$) covers values in a range of more than one order of magnitude. The horizontal shaded band is value of the Standard Model of cosmology.} \label{uncertainties}
 \end{center}
\end{figure}

For the allowed region in Fig. \ref{msugra1} discussed sofar the relic density constraint has not yet been considered and the interesting question is: does this rather narrow region yield the correct relic density? The answer is simple: almost any pair of $m_0$ and $m_{1/2}$ values are allowed by the relic density, if the values of the remaining parameters, namely $A_0$ and $\tan\beta$ are chosen accordingly. For large values of $\tan\beta$, typically above 50, this is the case, as shown in Fig. \ref{msugra2}. In Fig. \ref{uncertainties} the relic density is shown as a function of $\tan\beta$ for a small value of $m_{1/2}$ which is compatible with the ERGET excess. The calculated value of the relic density has a strong dependence on the SM parameters like $\alpha_s$, $m_t$ and $m_b$ and the corresponding uncertainty covers values in a region which spans over more than one order of magnitude. For large values of $\tan\beta$ the relic density is very sensitive to $\tan\beta$ and the SM parameters, which can be understood in the following way:

\begin{figure}
 \begin{center}
  \includegraphics [width=0.40\textwidth,clip]{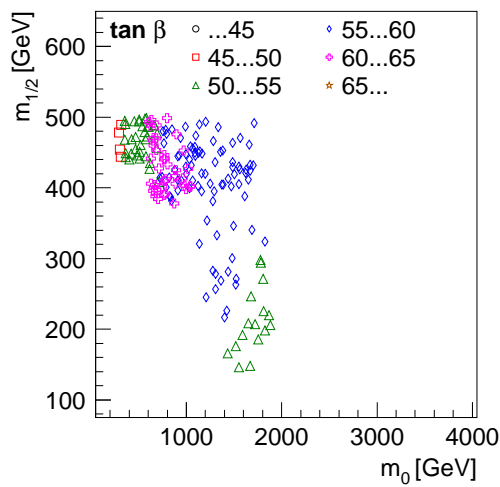}
  \caption[]{A random parameter scan has been performed, which checks for each point, if all experimental constraints, including the relic density, are fulfilled. One observes that the selected region of Fig. \ref{msugra1} requires values of $\tan\beta$ above 50, if the relic density constraint is to be fulfilled as well.} \label{msugra2}
 \end{center}
\end{figure}

In mSUGRA models the parameters are set at the GUT scale. The quantities like cross sections or masses have to be calculated at much lower energies, typically the electroweak scale. Therefore renormalization group equations are used to run the couplings and soft breaking terms from the unification scale to lower energies. This running also depends on the values of the SM parameters in particular on the Yukawa couplings of the third generation and the strong coupling constant:
\begin{eqnarray}
  {dm^2_{H_2}\over dt} &=& \ldots 
    - {3h_t^2\over (4\pi)^2} \left( \tilde{m}_Q^2+\tilde{m}_t^2+m_{H_2}^2+A_t^2m_0^2 \right) \nonumber\\
  {dm^2_{H_1}\over dt} &=& \ldots
    - {3h_b^2\over (4\pi)^2} \left( \tilde{m}_Q^2+\tilde{m}_b^2+m_{H_1}^2+A_b^2m_0^2 \right) \nonumber
\end{eqnarray}
with $t=\log(M_{\mbox{\tiny{GUT}}}^2/Q^2)$ and the initial conditions $m_{H_2}=m_{H_1}=m_0.$ The Yukawa couplings $h$ are determined by the running quark masses. For the running masses it is important to include higher order radiative corrections \cite{dmqSM} and supersymmetric contributions \cite{dmqMSSM}, since they are not negligible. In particular corrections to $m_b$ become more and more important with increasing $\tan\beta$. Since they are negative, the coupling is getting large and the RGE equation for $m_{H_1}$ is dominated by the corresponding $h_b^2$-term. So a difference in the measured quark mass is assigned to the mass of the pseudoscalar Higgs mass by the definition of $m_A$ (at tree level):
\begin{equation}
  m_A^2=m_1^2+m_2^2=m_{H_1}^2+m_{H_2}^2+2\mu^2 \nonumber
\end{equation}
Electroweak symmetry breaking is triggered if the following condition is fullfilled:
\begin{equation}
  {M_Z^2\over 2} = {m_1^2 - m_2^2\tan^2\beta \over \tan^2\beta-1} \approx -m_2^2 \nonumber
\end{equation}
The parameter $m_2$ depends on the RGE of $m_{H_2}$. For large values of $m_0$ the running of $m_{H_2}$ is very steep and sensitive to $m_t$ \cite{bs}. The reason for this is the large $h_t$ and the heavy sfermion masses, since $m_0$ is large, i.e. the term in the RGE of $m_{H_2}$ proportional to $h_t^2$ gets dominant. If $\tan\beta$ is small than $h_b$ is also small and the running of $m_{H_1}$ is not steep, since the term which includes the sfermion masses is small. In this case $m_{H_2}$ turns out to have a value determined by the condition above with a relatively small theoretical uncertainty, since $M_Z$, $\tan\beta$ and $m_1$ have small errors. 

In case of large $\tan\beta$ the running of $m_{H_1}$ starts to be steep too, due to the increasing bottom Yukawa coupling. The steep running of $m_{H_2}$ and $m_{H_1}$ has as an effect that from the minimization condition the parameters $m_2$ and $m_1$ cannot be determined with high accuracy since two of the quantities in the relation have large uncertainties. This causes a high sensitivity to the radiative corrections of the Higgs potential. Only a small change of these corrections or the heavy quark masses will result in very different values of $m_{H_1}$ and $m_{H_2}$ and correspondingly $m_A$ has large theoretical uncertainty.

For such large values of $\tan\beta$ the annihilation via pseudoscalar Higgs exchange, being proportional to $(\tan\beta)^2$, becomes dominant, if the neutralino mass is not close to half the $Z^0$ mass. In the latter case the annihilation is almost a pure $s$-wave, i.e. at low energies the cross section is suppressed. Therefore the fact that the annihilation is still observable in the present universe, as indicated by the EGRET excess, implies a lower limit of about 50 GeV on the neutralino mass, unless one allows for an extreme boosting of the DMA by the fact that the DM is clustered. One would need boost factors of several thousands \cite{us}.

A typical mass spectrum for the allowed region of Fig. \ref{msugra1} and $\tan\beta=52.2$ is given in Table \ref{t1}. It should be noted that the annihilation is still in the so-called bulk region, i.e the regions not dominated by co-annihilation or resonances, since the neutralino mass is far away from the pseudoscalar Higgs mass resonance for the mSUGRA spectrum and not close to any of the other sparticles, like stau or chargino, as shown in Table \ref{t1}. The total annihilation rate, which is the sum of the self-annihilation and co-annihilation rate, is fixed by the observed relic density. Therefore a large co-annihilation rate automatically implies a negligible self-annihilation rate. Since in the present universe the NLSPs have decayed, only the self-annihilation is operative now and would be practically zero in case of strong co-annihilation. So it is good for indirect DM detection that the combination of EGRET data with Higgs mass limits results in a spectrum, for which the co-annihilation is negligible.

\begin{table}
  \caption{Typical mSUGRA parameters from the EGRET analysis, sparticle masses and observables.}
  \label{t1}       
  \begin{tabular}{ll}
    \hline\noalign{\smallskip}
    Parameter & value \\
    \noalign{\smallskip}\hline\noalign{\smallskip}
    $m_0$ & 1500 GeV \\
    $m_{1/2}$ & 170 GeV \\
    $A_0$ & $0\cdot m_0$ \\
    $\tan\beta$ & 52.2 \\
    \noalign{\smallskip}\hline\noalign{\smallskip}
    $\alpha_s(M_Z)$ & 0.122 \\
    $m_t (pole)$ & 175 GeV\\
    $m_b(m_b)$ & 4.214 GeV\\
    \noalign{\smallskip}\hline\noalign{\smallskip}
    $\tilde \chi^0_{1,2,3,4}$ & 64, 113, 194, 229 \\
    $\tilde \chi^\pm_{1,2},\tilde{g}$ & 110, 230, 516 \\
    $\tilde t_{1,2}$ & 906, 1046 \\
    $\tilde b_{1,2}$ & 1039, 1152 \\
    $\tilde \tau_{1,2}$ & 1035, 1288 \\
    $\tilde \nu_e, \tilde \nu_\mu, \tilde \nu_\tau$ & 1495, 1495, 1286 \\
    $h,H,A,H^\pm$ & 115, 372, 372, 383 \\
    \noalign{\smallskip}\hline\noalign{\smallskip}
    $Br(b\to X_s\gamma)$ & $3.02 \cdot 10^{-4}$ \\
    $\Delta a_\mu$ & $1.07\cdot 10^{-9}$ \\
    $\Omega h^2$ & 0.117 \\
    \noalign{\smallskip}\hline
  \end{tabular}
\end{table}

\section{Conclusion}

In our previous paper \cite{us} the observed excess of diffuse Galactic gamma rays was shown to exhibit all the features of Dark Matter Annihilation, especially the spatial distribution of the excess was shown to trace the DM distribution, as proven by the fact that one could reconstruct the peculiar shape of the rotation curve of our Galaxy from the gamma ray excess. In this paper the DM interpretation of the EGRET excess is compared with Supersymmetry and it is shown that the minimal supersymmetric model with the popular supergravity inspired symmetry breaking, gauge unification and radiative electroweak symmetry breaking is in perfect agreement with the EGRET excess. The mass spectrum of the gauginos is governed by the neutralino mass corresponding to $m_{1/2}$ roughly between 125 and 175 GeV, while the scalar masses are constrained by the Higgs mass and/or $Br(B \to X_s\gamma)$ to have $m_0$ above 1200 GeV and to be below roughly 2 TeV in order to allow for EWSB. The upper limit has a large uncertainty from the Yukawa coupling of the third generation and the gauge couplings, since the steep running of the mass terms of the Higgs potential are governed by them. Such a mass spectrum is observable at the LHC. If confirmed, especially a neutralino mass around 60 GeV, then this would prove that DM can indeed be considered the supersymmetric partner of the Cosmic Microwave Background, since the neutralino is almost a pure bino in this case.

\end{document}